\documentclass[a4paper, 10pt, conference]{ieeeconf}      % Use this line for a4 paper

\IEEEoverridecommandlockouts                              % This command is only needed if 
\usepackage{cite}
\usepackage{amsmath,amssymb,amsfonts}
\usepackage{algorithmic}
\usepackage{graphicx}
\usepackage{textcomp}
\usepackage{xcolor}
\usepackage{glossaries}
\usepackage{tikz}
\usetikzlibrary{arrows,calc}
\usepackage{pgfplots}
\usepackage{siunitx}
\usepackage{url}
\usepgfplotslibrary{groupplots}
\usetikzlibrary{shapes.misc}
\tikzset{cross/.style={cross out, draw, 
		minimum size=2*(#1-\pgflinewidth), 
		inner sep=0pt, outer sep=0pt, very thick}}
\usetikzlibrary{arrows.meta}
\usetikzlibrary{calc,patterns,angles,quotes}

\usepackage{textcomp}
\usepackage{lipsum}
\newcommand\copyrighttext{%
	\footnotesize \textcopyright 2019 IEEE.  Personal use of this material is permitted.  Permission from IEEE must be obtained for all other uses, in any current or future media, including reprinting/republishing this material for advertising or promotional purposes, creating new collective works, for resale or redistribution to servers or lists, or reuse of any copyrighted component of this work in other works.
}
\newcommand\copyrightnotice{%
	\begin{tikzpicture}[remember picture,overlay]
	\node[anchor=south,yshift=10pt, xshift=10pt] at (current page.south) {\fbox{\parbox{\dimexpr\textwidth-\fboxsep-\fboxrule\relax}{\copyrighttext}}};
	\end{tikzpicture}%
}

\newacronym{EMS}{EMS}{Energy Management Strategy}
\newacronym{DC}{DC}{Direct Current}
\newacronym{AC}{AC}{Alternating Current}
\newacronym{OCP}{OCP}{Optimal Control Problem}
\newacronym{NLPP}{NLPP}{Nonlinear Programming Problem}
\newacronym{NLP}{NLP}{Nonlinear Programming}
\newacronym{MILP}{MILP}{Mixed Integer Linear Programming}
\newacronym{MIQP}{MIQP}{Mixed Integer Quadratic Programming}
\newacronym{OCV}{OCV}{Open Circuit Voltage}
\newacronym{COG}{COG}{Center of Gravity}
\newacronym{SOC}{SOC}{State of Charge}
\newacronym{TUM}{TUM}{Technical University of Munich}
\newacronym{ODEs}{ODEs}{Ordinary Differential Equations}

\title{\LARGE \bf
Energy Management Strategy for an Autonomous Electric Racecar using Optimal Control
}

\author{Thomas Herrmann$^{1}$, Fabian Christ$^{2}$, Johannes Betz$^{1}$ and Markus Lienkamp$^{1}$% <-this % stops a space
\thanks{$^{1}$Thomas Herrmann, Johannes Betz and Markus Lienkamp are with the Chair of Automotive Technology, Faculty of Mechanical Engineering, Technical University of Munich, 85748 Garching b. Muenchen, Germany
        {\tt\small thomas.herrmann@tum.de}}%
\thanks{$^{2}$Fabian Christ is with the Chair of Automatic Control, Department of Mechanical Engineering, Technical University of Munich, 85748 Garching b. Muenchen, Germany}%
}

\begin{document}

\maketitle
\copyrightnotice
\thispagestyle{empty}
\pagestyle{empty}

%%%%%%%%%%%%%%%%%%%%%%%%%%%%%%%%%%%%%%%%%%%%%%%%%%%%%%%%%%%%%%%%%%%%%%%%%%%%%%%%
\begin{abstract}
The automation of passenger vehicles is becoming more and more widespread, leading to full autonomy of cars within the next years. Furthermore, sustainable electric mobility is gaining in importance. As racecars have been a development platform for technology that has later also been transferred to passenger vehicles, a race format for autonomous electric racecars called \textit{Roborace} has been created.\\
As electric racecars only store a limited amount of energy, an \gls{EMS} is needed to work out the time as well as the minimum energy trajectories for the track. At the same time, the technical limitations and component behavior in the electric powertrain must be taken into account when calculating the race trajectories. In this paper, we present a concept for a special type of \gls{EMS}. This is based on the \gls{OCP} of generating a time-minimal global trajectory which is solved by the transcription via direct orthogonal collocation to a \gls{NLPP}. We extend this minimum lap time problem by adding our ideas for a holistic \gls{EMS}. This approach proves the fundamental feasibility of the stated ideas, e.g. varying race-paths and velocities due to energy limitations, covered by the \gls{EMS}. Also, the presented concept forms the basis for future work on meta-models of the powertrain's components that can be fed into the \gls{OCP} to increase the validity of the control output of the \gls{EMS}.
\end{abstract}

%%%%%%%%%%%%%%%%%%%%%%%%%%%%%%%%%%%%%%%%%%%%%%%%%%%%%%%%%%%%%%%%%%%%%%%%%%%%%%%%
\section{INTRODUCTION}

In 2018, the \gls{TUM} participated in the first Roborace event \cite{Roborace2018}. Roborace stages the first race series for autonomous vehicles (\textit{Robocars}) and is a support series for Formula E. The software stack developed by the team at TUM used to operate the Robocar \cite{Betz2019b} is already partially publicly available \cite{Wischnewski2019}. This paper presents a concept and the main ideas for an \gls{EMS} that extends the software module calculating the global trajectories leading to the minimum lap time \cite{Christ2019b}. The \gls{EMS} is crucial as it considers component behavior and the inherent limitations of the all-electric powertrain. This is necessary because of the significant influence of components on the minimum achievable race time in total over all laps.\\
According to \cite{Betz2019}, the results from research in the field of autonomous motorsport provide information on future autonomous road vehicles for the following three reasons:
\begin{itemize}
	\item The algorithms developed for and tested in an autonomous racecar must be capable of being calculated using limited computational resources in real-time and must be exceptionally robust.
	\item As the Robocars have electric powertrains, the development of an \gls{EMS} that enables energy-saving and energy recovery ensures progress beyond established technologies, including series production vehicles.
	\item Since it is also included in the presented \gls{EMS}, the choice of trajectory is of major interest for autonomous passenger vehicles as well as racecars for reasons of safety, range or a minimal lap time.
\end{itemize}
For these reasons, the \gls{EMS} is being developed in the context of autonomous motorsport for testing under extremely tough conditions in an enclosed environment, where technical effects and correlations of the powertrain components can be clearly seen.\\
The structure of this paper is as follows: In Section \ref{sec:stateofart}, the state of the art is being summarized. Section \ref{subsec:concept} contains the concept of the presented \gls{EMS}. Section \ref{sec:methodology} describes the methods used together with formulation of the \gls{OCP}, including its states and control inputs as well as the powertrain architecture of an electric rear wheel drive vehicle. The results of the presented \gls{OCP} are described in Section \ref{sec:results}, Section \ref{sec:conclusion} summarizes the results obtained and also states the direction of future work and how the presented \gls{OCP} will be extended.

\section{State of the Art}
\label{sec:stateofart}
The optimal control of vehicles is a complex field that has been dealt with in prior publications. In the following, several different mathematical approaches are summarized in order to plan velocities, paths or entire trajectories with optimal control based approaches. Their specific advantages and drawbacks are described.\\
Within this section we do not distinguish between the specific objectives that are being optimized in these approaches. They are classified according to the mathematical optimization methods, to solve the \gls{OCP}.
	\subsubsection{Explicit solution}
	The authors of \cite{Dib2014} and \cite{Sciarretta2015} calculate one-dimensional velocity profiles for predefined paths to minimize the energy demand of the traveling vehicle. The vehicle's dynamics are expressed using a simplified point mass model \cite{Petit2011}. The driving resistance is described applying Newton's second law resulting in
	\begin{equation}
		m\ddot{x} = \frac{1}{r} M_\mathrm{e} i_\mathrm{g} - \frac{1}{2}\rho_\mathrm{a} A c_\mathrm{w} \dot{x}^2 - mgc_\mathrm{r} -mg\sin{\left(\alpha(x)\right)}
		\label{eq:Newton}
	\end{equation}
	where $x$ is the vehicle's position, $m$ the vehicle mass, $M_\mathrm{e}$ the output torque of the electric machine, $i_\mathrm{g}$ the gear transmission, $r$ the wheel radius, $\rho_\mathrm{a}$ the air density, $A$ the vehicle's front surface, $c_\mathrm{w}$ the aerodynamic drag coefficient, $g$ the gravitational acceleration, $c_\mathrm{r}$ the rolling resistance coefficient and $\alpha(x)$ the road slope.\\
	To bring the energy consumption $E_{\mathrm{\Sigma}}$ into play, \cite{Dib2014} introduces a second order polynomial of the form
	\begin{equation}
		E_{\mathrm{\Sigma}} = \int{b_\mathrm{1} \tau + b_\mathrm{2} \dot{x}(t) \tau^2 \mathrm{d}t}
	\end{equation}
	with $b_\mathrm{1}$, $b_\mathrm{2}$ being constant fitting coefficients and $\tau$ the required torque. Neglecting aerodynamic drag, \cite{Sciarretta2015} approximates the energy demand by
	\begin{equation}
		E_{\mathrm{\Sigma}} = \int{F_\mathrm{w} \dot{x}(t) \mathrm{d}t}
	\end{equation}
	where $F_\mathrm{w}$ depicts the traction force deduced from (\ref{eq:Newton}). The Hamiltonian is devised to deduce an explicit solution to the stated \gls{OCP}. As one can distinguish, only simple model equations can be formulated to be able to determine a solution using an explicit approach.

\subsubsection{Mixed Integer Linear/Quadratic Programming}
\gls{MILP} or \gls{MIQP} is used widely in order to generate optimal trajectories for specific driving maneuvers. In \cite{Qian2016}, an \gls{MIQP} formulation is used to solve problems like vehicle overtaking, obstacle avoidance or lane changes. To plan global trajectories, \cite{Ademoye2006} uses \gls{MILP} to find the time-minimal path while simultaneously avoiding static obstacles. Furthermore, \cite{Gorelik2018} uses \gls{MILP} to calculate the optimal energy distribution within an electric powertrain during driving to control fail-operational power nets. However, the driving kinematics modeled within the enumerated publications are based on simplified point mass models.

\subsubsection{Graph search}
To account for more complex models of vehicle dynamics, nonlinearities in objective functions or in the boundary conditions of the optimization problem, graph search approaches are used widely. The architecture used in \cite{Frazzoli2002} addresses the dynamic constraints on the vehicle’s motion in real-time. An RRT*-algorithm is implemented to solve the minimum lap-time problem using a half-car dynamic model to find its local steering input in \cite{Jeon2013}.

\subsubsection{Convex optimization}
There are several approaches to relax an \gls{OCP} of electric vehicles to a convex optimization problem \cite{Murgovski2015}, \cite{Ebbesen2018}. The main advantage of these re-formulations is the almost negligible calculation time needed to solve the defined problem. On the other hand, these problem formulations suffer from subsequent extension, since any additional equality or inequality constraint must also be formulated to fit into the existing framework. Furthermore, \cite{Ebbesen2018} and\cite{Salazar2018} assume a fixed driving path as well as a point mass to reduce the complexity of the optimization problem. This in turn enables calculation of a time-optimal velocity profile exploiting the convexity of the problem formulation.
%Publication \cite{Murgovski2015} assumes the \gls{OCV} of the battery to be constant. But a variable \gls{OCV} is of major interest when describing the behavior of an electric powertrain in highly dynamic driving scenarios.
\subsubsection{Nonlinear Programming}
\cite{Casanova2000} uses \gls{NLP} to solve a minimal lap time problem for a Formula 1 racecar modeled as nonlinear double track model including a detailed tire model. The problem formulation is extended in \cite{Kirches2010} taking gear shifts into account. Track-specific parameter optimization is done in \cite{ImaniMasouleh2016}. \cite{Limebeer2015} also considers three-dimensional track courses when solving the minimal lap time problem. As one can recognize, complex dynamic scenarios are modeled using an \gls{NLP} approach. Unfortunately, the computing times for solving these are relatively long in comparison with other mathematical problem formulations.
\section{Concept}
\label{subsec:concept}
\begin{figure}[h]
	\centering
	\definecolor{colBlau}{RGB}{0,101,189}%
\definecolor{colBlueDark}{RGB}{0,82,147}%
\definecolor{colRed}{RGB}{227,114,34}%
\definecolor{colGreen}{RGB}{162,173,0}%
\definecolor{colGray}{RGB}{153,153,153}%
\tikzstyle{simpleNode} = [rectangle, rounded corners, minimum width=1cm, minimum height=1cm,text centered, draw=black, line width=0.5mm]
\tikzstyle{arrow} = [->,-triangle 45,line width=0.5mm,black]
\tikzstyle{lineE} = [-,line width=0.5mm,colGray,dashed]
\tikzstyle{lineM} = [-,line width=1.5mm,black]
\begin{tikzpicture}[node distance=1cm]
\coordinate (Zero) at (0,0);
\node (Battery) [simpleNode, align=center] {$B$};
\node (InverterF) [simpleNode, align=center, left of=Battery, xshift=-0.5cm] {$I_{\mathrm{l/r}}$};
\node (MotorF) [simpleNode, align=center, left of=InverterF,xshift=-0.5cm] {$M_{\mathrm{l/r}}$};
\node (GearF) [simpleNode, align=center, left of=MotorF, xshift=-0.15cm] {$G_{\mathrm{l/r}}$};
\node (WheelFR) [simpleNode, align=center, above of=GearF, minimum height=0.6cm, yshift=0.3cm] {$W_{\mathrm{rl}}$};
\node (WheelFL) [simpleNode, align=center, below of=GearF, minimum height=0.6cm, yshift=-0.3cm] {$W_{\mathrm{rr}}$};
\node (Aux) [simpleNode, align=center, above of=Battery, yshift=0.3cm] {$A_\mathrm{x}$};
\node (InverterR) [simpleNode, minimum width=0cm, right of=Battery, draw=white, xshift=0.25cm] {};
%%% Edges %%%
\draw [lineE] ([yshift=0.2cm]Battery.west) to ([yshift=0.2cm]InverterF.east);
\draw [lineE] ([yshift=-0.2cm]Battery.west) to ([yshift=-0.2cm]InverterF.east);
\draw [lineE] ([yshift=0.2cm]Battery.east) to ([yshift=0.2cm]InverterR.west);
\draw [lineE] ([yshift=-0.2cm]Battery.east) to ([yshift=-0.2cm]InverterR.west);
% To Aux
\draw [lineE] ([xshift=0.2cm]Battery.north) to ([xshift=0.2cm]Aux.south);
\draw [lineE] ([xshift=-0.2cm]Battery.north) to ([xshift=-0.2cm]Aux.south);
\draw [lineE] (InverterF)[anchor=left] to (MotorF)[anchor=right];
\draw [lineE] ([yshift=0.25cm]InverterF.west) to ([yshift=0.25cm]MotorF.east);
\draw [lineE] ([yshift=-0.25cm]InverterF.west) to ([yshift=-0.25cm]MotorF.east);
\draw [lineM] (MotorF)[anchor=left] to (GearF)[anchor=right];
\draw [lineM] (GearF)[anchor=left] to (WheelFR)[anchor=right];
\draw [lineM] (GearF)[anchor=left] to (WheelFL)[anchor=right];
%\draw [arrow] (Battery) to[bend left] (InverterF);
\draw [thick] ($(InverterR.west)+(0,1)$) to [out=300,in=120] ($(InverterR.west)+(0,-1)$);
\end{tikzpicture}
	\caption{Electric powertrain architecture of a rear wheel drive vehicle}
	\label{fig:PowertrainRobocar}
\end{figure}
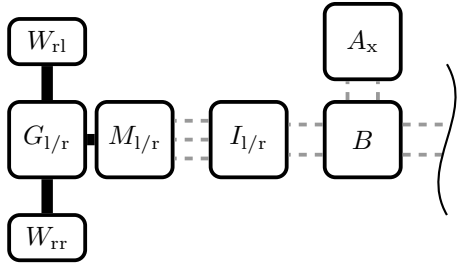

This paper covers the concept and the main ideas for an \gls{EMS} for autonomous electric cars. This \gls{EMS} aims to find the minimum race time by optimizing global lap trajectories (path \& velocity) while taking the technical constraints of the components in the all-electric powertrain into account. The results in this paper demonstrate the overall feasibility of the stated \gls{OCP} for one lap trajectory. The formulation of this optimal control based approach enables us to extend the problem formulation by component behaviors in the future. Furthermore, we will be able to consider multiple consecutive race laps.\\
On the racetrack, highly dynamic driving scenarios, including maximum velocities and peak positive as well as negative accelerations, occur. These put enormous stress on electric powertrain components. Additionally, environmental conditions vary, depending on race locations. Therefore, extreme heat or cold or humidity and aridness can occur. Due to these facts, component behavior must be considered when solving the minimal race time \gls{OCP}.\\
Our approach is based on our previous work, as presented in \cite{Christ2019b}. Therein, an \gls{OCP} for planning time-optimal trajectories was formulated that allows for easy subsequent extension via the components' behavior within the electric powertrain. In this way, we can consider technical constraints and include energetic considerations in the \gls{OCP}. This extension enables planning of the global race trajectories for all race laps that need to be completed.\\
Effects within the powertrain leading to unexpected component behavior or limited available power can be due to:
\begin{itemize}
	\item Reaching the maximum permitted battery temperature.
	\item The maximum permitted temperature of the electric machines are reached, especially during qualifying, as higher peak power is allowed there compared to the race itself.
	\item A decreased level of efficiency of the motor inverters due to the battery's voltage drop to equal the input voltage of the motor inverters and due to increased inverter temperature.
	\item Quadratically higher thermal power loss ($P_{\mathrm{l,T}} \propto I^2$) due to higher current $I$ within the powertrain owing to the continuous voltage decrease in accordance with the \gls{SOC} of the main battery.
\end{itemize}
The most important components of the electric powertrain with rear wheel drive are depicted in Fig. \ref{fig:PowertrainRobocar} with
\begin{itemize}
	\item Energy storage represented as the battery ($B$).
	\item Power electronics converting the battery's \gls{DC} into \gls{AC} at rear left ($I_\mathrm{l}$) and rear right ($I_\mathrm{r}$).
	\item Synchronous permanent electric machines at the rear left ($M_\mathrm{l}$) and rear right ($M_\mathrm{r}$).
	\item Gears attached to the electric machines ($G_\mathrm{{l/r}}$) to transform the motor torque into drive torque.
	\item Sensors for autonomous driving ($A_\mathrm{x}$) that need to be powered by the battery and must not be neglected, since they consume a major amount of the vehicle's total energy demand. 
\end{itemize}
The wheels are $W_\mathrm{{rl}}$ and $W_\mathrm{{rr}}$; displayed is only the rear part of the whole powertrain.

\section{METHODOLOGY}
\label{sec:methodology}
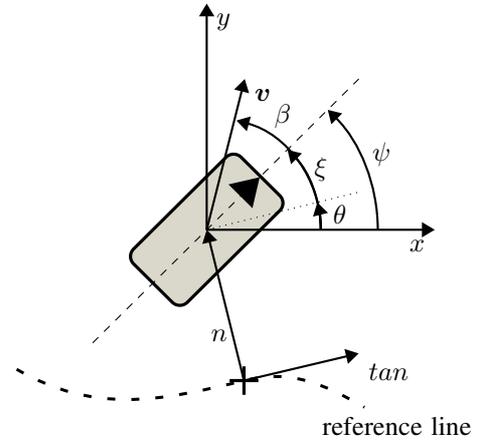
\begin{figure}[h]
	\centering
	\definecolor{colBlau}{RGB}{0,101,189}%
\definecolor{colBlueDark}{RGB}{0,82,147}%
\definecolor{colRed}{RGB}{227,114,34}%
\definecolor{colGreen}{RGB}{162,173,0}%
\definecolor{colGray}{RGB}{218,215,203}%
\tikzstyle{simpleNode} = [rectangle, rounded corners, minimum width=1cm, minimum height=1cm,text centered, draw=black, line width=0.5mm]
\tikzstyle{arrow} = [->,-triangle 45,line width=0.5mm,black]
\tikzstyle{lineE} = [-,line width=0.5mm,colGray,dashed]
\tikzstyle{lineM} = [-,line width=1.5mm,black]
\begin{tikzpicture}[node distance=1cm]
\coordinate (Zero) at (0,0);
\coordinate (ZeroS) at (0.49,-2);
\draw[very thick, rotate=-45, rounded corners, fill=colGray] (-0.5,-1) rectangle (0.5,1);
\draw[line width=0.9mm,-Triangle] (0.5,0.5) -- (0.7,0.7);
\draw[thick,-Triangle] (0,0) -- (3,0) node[anchor=north east]{$x$};
\draw[thick,-Triangle] (0,0) -- (0,3) node[anchor=north west]{$y$};
\draw[dashed] (-1.5,-1.5) -- (2,2) node[anchor=north west]{};
\draw[dotted] (0,0) -- (2,0.5) node[anchor=north west]{};
\draw[thick,-Triangle] (ZeroS) -- (Zero) node[anchor=north east,yshift=-1.2cm,xshift=0.4cm]{$n$};
\draw[thick,-Triangle] (ZeroS) -- (2,-1.64) node[anchor=north west]{$tan$};
\node[](theta) at (1.75,0.2) {$\theta$};
\node[](xi) at (1.5,0.8) {$\xi$};
\node[](beta) at (1,1.5) {$\beta$};
\node[](psi) at (2.3,1) {$\psi$};
\draw[thick,-Triangle] (0,0) -- (0.5,2) node[anchor=north west]{$\boldsymbol{v}$};
\draw[thick,-Triangle] (1.5,0) arc (0:14.04:1.5);
\draw[thick,-Triangle] (1.5,0) arc (0:45:1.5);
\draw[thick,-Triangle] (2.25,0) arc (0:45:2.25);
\draw[thick,-Triangle] (1.5,0) arc (0:74.96:1.5);
\draw (ZeroS) node[cross=5pt,rotate=45]{};
\node[](refEnd) at (2.5,-2.6) {reference line};
\draw [line width=0.4mm, loosely dashed] (-2.5,-1.85) [out=330,in=194.04] to (ZeroS) to [out=14.04,in=150] (refEnd);
\end{tikzpicture}
	\caption{Vehicle and path model used in the \gls{OCP} \cite{Christ2019b}}
	\label{fig:states_plot}
\end{figure}

Section \ref{subsec:concept} presented a concept and the main ideas for the \gls{EMS} for an autonomous electric racecar. This strategy will lead to the minimum race time $T_{\mathrm{\Sigma}}$ totaled over all the race lap times~$T_i$,
\begin{equation}
	T_{\mathrm{\Sigma}} = \sum_{i=1}{T_i}.
\end{equation}
The vehicle's dynamics are described using a nonlinear double track model that includes longitudinal, lateral and yaw freedoms. This means that a quasi-steady state wheel load transfer is permitted whenever the car is accelerating or cornering. For a detailed description of the required first-order ordinary differential equations describing the model dynamics, we refer to~\cite{Christ2019b}.\\
\\
An \gls{OCP} including equality and inequality constraints to be solved by the \gls{EMS} is defined by \cite[p.~478]{Bertsekas2016}, \cite[p.~127]{Boyd2004}, \cite[p.~215]{Sioshansi2017}:
\begin{align}	
	\min~& l(\boldsymbol{x})\\
	s.t.~\frac{\mathrm{d}\boldsymbol{x}}{\mathrm{d}s} &= f(\boldsymbol{x}(s), \boldsymbol{u}(s))\\
	h_\mathrm{i} &= 0\\
	g_\mathrm{j} &\leq 0
	\label{eq:NLPP}
\end{align}
with $i = 1, ..., m$ and $j = 1, ..., r$, where $s$ as the independent variable within the \gls{OCP} denotes the distance along the reference line of the racetrack (Fig.~\ref{fig:states_plot}).\\
The following summarizes the formulation of the time-minimal \gls{OCP} defined in our previous work \cite{Christ2019b} and is extended by energy-related considerations.\\
The state vector $\boldsymbol{x}(s)$ within the \gls{OCP} is defined as
\begin{equation}
	\boldsymbol{x}(s) = 
	\begin{pmatrix}
	v~\beta~\dot{\psi}~n~\xi
	\end{pmatrix}^T
\end{equation}
with the vehicle's states $v$, $\beta$ as the side slip angle and $\psi$ the yaw angle of the vehicle as well as the path model's states $n$ denoting the lateral distance to the reference line and $\xi$ being the relative angle of the vehicle's longitudinal axis to the tangent $tan$ on the reference line. Fig. \ref{fig:states_plot} visualizes the used states of the \gls{OCP}. $\theta$ denotes the angle between tangent and local $x$-axis. The gray rectangle with rounded corners represents the vehicle heading north-east.
\\
The control input vector $\boldsymbol{u}(s)$ is defined by
\begin{equation}
\boldsymbol{u}(s) = 
\begin{pmatrix}
	F_\mathrm{d}~F_\mathrm{b}~\delta~\gamma
\end{pmatrix}^T
\end{equation}
with $F_\mathrm{d}$ being the driving force, $F_\mathrm{b}$ the braking force, $\delta$ the steering input and $\gamma$ the wheel load transfer according to the nonlinear double track model \cite{Christ2019b}.\\
The independent variable $s$ changes according to
\begin{equation}
	\dot{s} = \frac{v\cos{\left(\xi + \beta\right)}}{1 - n\kappa}.
\end{equation}
The transitions of the states in $\boldsymbol{x}(s)$ can be summarized as follows:
\begin{align}
\dot{v} &= \pi_\mathrm{1}(v,\beta,\delta,F_{\mathrm{tire}})
\label{eq:dot_v}\\
\dot{\beta} &= \pi_\mathrm{2}(v,\beta,\delta,\dot{\psi},F_{\mathrm{tire}})\\
\ddot{\psi} &= \pi_\mathrm{3}(\delta,F_{\mathrm{tire}})
\label{eq:ddot_psi}\\
\dot{n} &= v \sin{\left(\xi + \beta\right)}
\label{eq:dot_n}\\
\dot{\xi} &= \dot{\psi} - \kappa\frac{v\cos{\left(\xi + \beta\right)}}{1 - n\kappa}
\label{eq:dot_xi}
\end{align}
where $\kappa = \frac{1}{R}$ describes the geometrical curvature by the inverse radius $R$ of a local curve in the reference line. The $\pi_\mathrm{i} (\cdot)$ denote a mathematical function. For a detailed description of these equations, we refer to our previous work \cite{Christ2019b}.

Here, $f(\boldsymbol{x}(s), \boldsymbol{u}(s))$ describes the system dynamics and contains (\ref{eq:dot_v}) - (\ref{eq:dot_xi}), as well as a functional correlation of the wheel load transfers according to the used nonlinear double track model \cite{Christ2019b}.\\
The objective function $l(\boldsymbol{x})$ that minimizes the race time can be written as
\begin{equation}	
	l(\boldsymbol{x}) = \int_{0}^{S_{\mathrm{\Sigma}}}{L(\boldsymbol{x}(s),\boldsymbol{u}(s)) \mathrm{d}s}
	\label{eq:objective}
\end{equation}
with $L(\cdot)$ being the Langrangian cost function defined in \cite{Gundlach2017}
\begin{equation}
	L = \frac{\mathrm{d}t}{\mathrm{d}s} = \frac{1 - n\kappa}{v\cos{\left(\xi + \beta\right)}}
\end{equation}
and $S_{\mathrm{\Sigma}}$ the entire race distance cumulated over all race laps.\\
The Lagrangian is also needed to transform the time-dependent \gls{ODEs} into space-dependent ones.\\ 
The equality constraints are
\begin{align}
	h_\mathrm{1} &= \gamma - \Pi\\
	h_\mathrm{2} &= F_\mathrm{d}\cdot F_\mathrm{b}
\end{align}
with $\Pi$ denoting the lateral wheel load transfer. The inequality constraints are
\begin{align}
	g_1 &= \sqrt{F_\mathrm{x}^2 + F_\mathrm{y}^2} - 1\\
	g_2 &= F_\mathrm{d}v -P_{\mathrm{max}}\\
	g_3 &= F_\mathrm{d} - F_{\mathrm{d,max}}\\
	g_4 &= - F_\mathrm{d}\\
	g_5 &= F_\mathrm{b}\\
	g_6 &= \delta - \delta_{\mathrm{max}}\\
	g_7 &= \delta_{\mathrm{min}} - \delta\\
	g_8 &= \frac{\Delta F_\mathrm{d}}{L\Delta s} - \frac{F_{\mathrm{d,max}}}{T_\mathrm{d}}\\
	g_9 &= \frac{F_{\mathrm{b,max}}}{T_\mathrm{b}} - \frac{\Delta F_\mathrm{b}}{L\Delta s}\\
	g_{10} &= \frac{\Delta \delta}{L\Delta s} - \frac{\delta_{\mathrm{max}}}{T_{\mathrm{\delta}}}\\
	g_{11} &= \frac{\delta_{\mathrm{min}}}{T_{\mathrm{\delta}}} - \frac{\Delta \delta}{L\Delta s},
\end{align}
where $F_\mathrm{x}$ and $F_\mathrm{y}$ are the longitudinal and the lateral forces set down by the tire. $P_\mathrm{{max}}$ is the maximum traction power of the vehicle, $F_{\mathrm{d,max}}$ and $F_{\mathrm{d,min}}$ describe the maximum drive and break force, respectively, $\delta_{\mathrm{max}}$ and $\delta_{\mathrm{min}}$ denote the maximum positive and the minimal negative steering angle input. With $T_\mathrm{j}$, appropriate time constants are defined to restrict the actuator dynamics \cite{Gundlach2017}.\\
The energy consumption during driving $E_{\mathrm{\Sigma}}$ is calculated using
\begin{align}
	E_{\mathrm{\Sigma}} &= \int_{0}^{T_{\mathrm{\Sigma}}}{P_\mathrm{d} + \bar{f}_\mathrm{r} P_\mathrm{b} \mathrm{d}t}\\
	&= \int_{0}^{T_{\mathrm{\Sigma}}}{(F_\mathrm{d} + \bar{f}_\mathrm{r} F_\mathrm{b})v \mathrm{d}t}
\end{align}
with $P_\mathrm{d}$ being the driving power, $P_\mathrm{b}$ the braking power and $\bar{f}_\mathrm{r}$ denoting a mean recuperation factor between \SI{0}{\percent} and \SI{100}{\percent}. The braking energy can then partially be re-stored in the battery (Fig. \ref{fig:PowertrainRobocar}).\\
The inequality constraint
\begin{align}
	g_{12} &= E_{\mathrm{\Sigma}} - \bar{E}_{\mathrm{\Sigma}}
\end{align}
results, limiting the maximum available amount of energy ($\bar{E}_{\mathrm{\Sigma}}$) for the entire race.
%
%
%%%%%%%%%%%%%%%%%%%%%%%%%%%%%%%%%%%%%%%%%%%%%%%%%%%%%%%%%%%%%%%%%%%%%%%%%%%%%%%%%%%%%%%%%%%%%%%%%%%%%%%%%%%%%%%%
\section{Results}
\label{sec:results}
\begin{figure}[h]
	%\begin{minipage}[b]{1\paperwidth}
	\centering
	\input{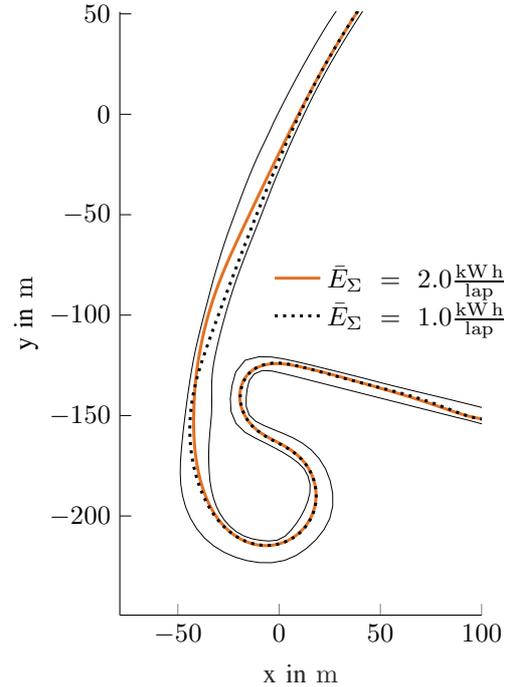}
	\caption{Time optimal raceline paths for different energy limitations on the Formula E racetrack in Berlin, Germany}
	\label{fig:paths}
	%\end{minipage}
\end{figure}

\begin{figure}[!h]
	%\begin{minipage}[b]{1\paperwidth}
	\input{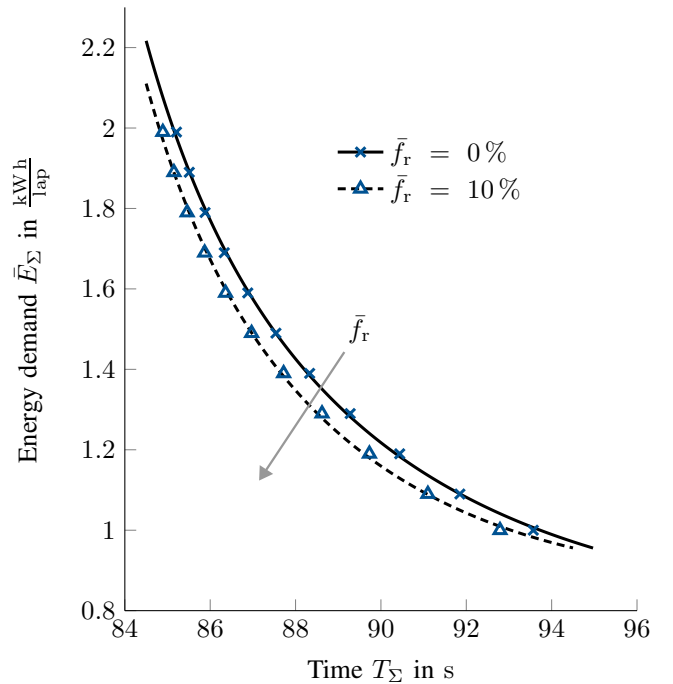}
	\caption{Pareto fronts containing the minimum reachable lap time $T_{\mathrm{\Sigma}}$ versus a limited energy demand $\bar{E}_{\mathrm{\Sigma}}$ including an average recuperation factor $\bar{f}_\mathrm{r}$}
	\label{fig:pareto_fronts}
	%\end{minipage}
\end{figure}

This section presents the results obtained with the primal-dual interior-point method IPOPT called by CasADi \cite{AnderssonInPress2018}. The execution time of the solver for the \gls{NLP} for one parameter set was always less than a minute on a PC with an i7-7820HQ CPU and 16GB of memory. The discretization step size is $\Delta s = \SI{5.0}{\meter}$.\\
In all of the following plots, the states $\boldsymbol{x}(s)$ as well as the control inputs $\boldsymbol{u}(s)$ are optimized to obtain the minimum race time. To show the feasibility of the presented concept, the race consists of a single lap on the Formula E track in Berlin, Germany.
\begin{figure}[h]
	%\begin{minipage}[b]{1\paperwidth}
	\input{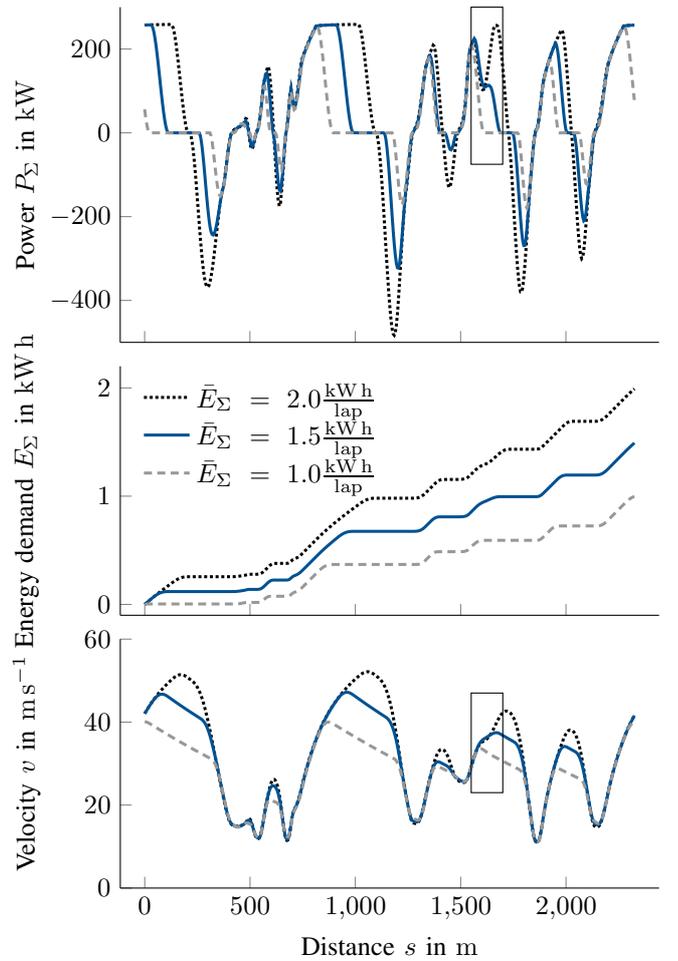}
	\caption{Power $P_{\mathrm{\Sigma}}$, energy demand $E_{\Sigma}$ and velocity $v(s)$ over the raceline distance $s$ on the Berlin Formula E racetrack}
	\label{fig:powEnVelDist}
	%\end{minipage}
\end{figure}\\
%\begin{figure}[h]
%	%\begin{minipage}[b]{1\paperwidth}
%	\input{Ressources/data_opt/pareto/pareto_ET_fit.tex}
%	\caption{Pareto fronts containing the minimal reachable lap time $T_{\Sigma}$ versus a limited energy demand including an average recuperation factor $\bar{f}_r$}
%	\label{fig:pareto_fronts}
%	%\end{minipage}
%\end{figure}
In Fig. \ref{fig:powEnVelDist}, the effects of different limitations of the maximum allowed amount of energy per lap $\bar{E}_{\mathrm{\Sigma}}$ are shown. The influence of $\bar{E}_{\mathrm{\Sigma}}$ can clearly be seen in a constantly repeating pattern: Peak velocity values are avoided if $\bar{E}_{\mathrm{\Sigma}}$ gets decreased due to the driving resistance growing quadratically with the velocity ($E_{\mathrm{\Sigma}} \propto v^2$). Peak power demands $P_{\mathrm{\Sigma}}(s)$ are requested on shorter distances and coasting phases ($P_{\mathrm{\Sigma}}(s) = \SI{0}{\kilo\watt}$) increase with a smaller $E_{\mathrm{\Sigma}}$. The behavior at $s\approx\SI{1600}{\meter}$ differs a little from the described pattern but reinforces this explanation (marked with rectangles in the plots): At this position on the racetrack, a curve to the right follows a short straight part. With the limited available amount of energy of $\bar{E}_{\mathrm{\Sigma}} = 1.0\frac{\SI{}{\kilo\watt\hour}}{\mathrm{lap}}$ no positive acceleration takes place before the curve. In contrast, with twice the amount of energy allowed for the lap $\bar{E}_{\mathrm{\Sigma}} = 2.0\frac{\SI{}{\kilo\watt\hour}}{\mathrm{lap}}$, a positive acceleration occurs resulting in a higher velocity on the straight part. These described patterns are similar to the technique of "lift and coast" in Formula E: Shortly before a curve, the optimized controlling policy suggests reduction of the peak power request and keeping the vehicle rolling for a short distance $\Delta s$. This behavior helps to lose kinetic energy without braking while simultaneously also reducing total energy demand.\\
In the middle of Fig. \ref{fig:powEnVelDist}, the cumulated energy demand $\bar{E}_{\mathrm{\Sigma}}$ is depicted. As the recuperation factor $f_\mathrm{r}$ was set to \SI{0}{\percent} for these experiments, these graphs either ascend or remain constant.
%\begin{figure}[h]
%	%\begin{minipage}[b]{1\paperwidth}
%	\input{Ressources/data_opt/power_energy_velocity_dist__all.tex}
%	\caption{Power $P_{\Sigma}$, energy demand $E_{\Sigma}$ and velocity $v(s)$ over the raceline distance $s$ on the Berlin Formula E racetrack}
%	\label{fig:powEnVelDist}
%	%\end{minipage}
%\end{figure}
As already stated in the concept in Section \ref{subsec:concept}, the time minimal path of the raceline varies, depending on the limited amount of every $\bar{E}_{\mathrm{\Sigma}}$ available per lap. In Fig.~\ref{fig:paths}, both displayed paths vary, especially from $(\SI{+25}{\meter},\SI{0}{\meter})$ to $(\SI{-40}{\meter},\SI{-200}{\meter})$ corresponding to the interval of $s=\left[\SI{0}{\meter},\SI{500}{\meter}\right]$ of the path distance. In this part of the racetrack the time-minimal path ($\bar{E}_{\mathrm{\Sigma}} = 2.0\frac{\SI{}{\kilo\watt\hour}}{\mathrm{lap}}$) is a little longer than the one for $\bar{E}_{\mathrm{\Sigma}} = 1.0\frac{\SI{}{\kilo\watt\hour}}{\mathrm{lap}}$. This is because of the lower curvature $\kappa$ following $s_{\mathrm{\bar{E}_{\Sigma}} = 2.0\frac{\SI{}{\kilo\watt\hour}}{\mathrm{lap}}}$, which simultaneously allows a higher peak velocity and therefore a smaller lap time.
\\
A summary of the results presented is given in Fig. \ref{fig:pareto_fronts}. Two Pareto fronts containing the correlation between the minimal reachable lap time $T_{\mathrm{\Sigma}}$ with a limited amount of available energy $\bar{E}_{\mathrm{\Sigma}}$ including an average recuperation factor~$\bar{f}_\mathrm{r}$ are shown. The hyperbolic form of the two fitted curves expresses the following: reducing the lap time leads to a squared increase in the energy demand $E_{\mathrm{\Sigma}}$ for the same lap. When introducing an average recuperation factor, converting negative braking force $F_\mathrm{b}$ to charge the battery (Fig. \ref{fig:PowertrainRobocar}), the Pareto front bends into the direction of the origin of the plot. With the help of these Pareto fronts, the decision on how to set up the race strategy can be made: the effects of a fast lap on energy consumption can be distinguished and the consequences for subsequent laps can be deduced. The future integration of powertrain components' behavior into the \gls{OCP}, as well as extension of the optimization problem to multiple race laps, help to determine a realistic race strategy for the entire race.
%\begin{table}[h]
%\caption{An Example of a Table}
%\label{table_example}
%\begin{center}
%\begin{tabular}{|c||c|}
%\hline
%One & Two\\
%\hline
%Three & Four\\
%\hline
%\end{tabular}
%\end{center}
%\end{table}
%%%%%%%%%%%%%%%%%%%%%%%%%%%%%%%%%%%%%%%%%%%%%%%%%%%%%%%%%%%%%%%%%%%%%%%%%%%%%%%%%%%%%%%%%%%%%%%%%%%%%%%%%%%%%%%
\addtolength{\textheight}{-0.5cm}
\section{CONCLUSION \& OUTLOOK}
\label{sec:conclusion}
In this paper, we presented an optimal control based approach for determining time-optimal lap trajectories for a rear wheel drive all-electric autonomous racecar, taking energy limitations into account. The solver IPOPT was able to find a feasible solution to the formulated \gls{OCP} in less than one minute on a standard user PC. The results show that the \gls{EMS} can determine the relationships between requested power, energy demand and path velocity when energy restrictions need to be met. Furthermore, the assumption that the optimal driving path differs from the time-minimal path when energetic limitations are considered, is confirmed.\\
This paper defines the basis for future work: The \gls{OCP} presented will be extended to multiple race laps instead of one. Furthermore, the behavior of the powertrain's components will be described using meta-models. These can then be included in the \gls{OCP} to achieve results taking the powertrain's states, e.g. temperatures of the energy storage or the electric machines, into account. With the help of these meta-models, a complete race strategy for a given powertrain configuration and an entire race can be determined.

%\addtolength{\textheight}{-20cm}   % This command serves to balance the column lengths

%%%%%%%%%%%%%%%%%%%%%%%%%%%%%%%%%%%%%%%%%%%%%%%%%%%%%%%%%%%%%%%%%%%%%%%%%%%%%%%%

%%%%%%%%%%%%%%%%%%%%%%%%%%%%%%%%%%%%%%%%%%%%%%%%%%%%%%%%%%%%%%%%%%%%%%%%%%%%%%%%

%%%%%%%%%%%%%%%%%%%%%%%%%%%%%%%%%%%%%%%%%%%%%%%%%%%%%%%%%%%%%%%%%%%%%%%%%%%%%%%%
\section*{CONTRIBUTIONS}

Thomas Herrmann initiated the idea of the paper and contributed significantly to the concept and the presented \gls{EMS}. Fabian Christ drew up the formulation of the  minimum lap-time problem as an \gls{OCP} including the numerical solver in his Master's thesis. Johannes Betz contributed to the whole concept of the paper. Markus Lienkamp provided a significant contribution to the concept of the research project. He revised the paper critically for important intellectual content. Markus Lienkamp gave final approval for the publication of this version and is in agreement with all aspects of the work. As a guarantor, he accepts responsibility for the overall integrity of this paper.

\section*{ACKNOWLEDGMENT}

We would like to thank the Roborace team for giving us the opportunity to work with them and for the use of their vehicles for our research project. We would also like to thank the \textit{Bavarian Research Foundation (Bayerische Forschungsstifung)} for funding us in connection with the "rAIcing" research project.\\
We would also like to thank Alexander Wischnewski and Leonhard Hermansdorfer for valuable discussions about the basic ideas behind this paper.
%\addtolength{\textheight}{20cm}

%%%%%%%%%%%%%%%%%%%%%%%%%%%%%%%%%%%%%%%%%%%%%%%%%%%%%%%%%%%%%%%%%%%%%%%%%%%%%%%%
%\dobeforekey{}{\newpage}
\bibliographystyle{IEEEtran}
\bibliography{IEEEabrv,referencesBIBTEX}

\end{document}